\documentclass[floatfix,onecolumn,aps,prd,amsmath,amssymb]{revtex4}
\usepackage{epsfig}

\usepackage[latin1]{inputenc}
\begin{document}

\title{Computer algebra systems as tools for learning Physics and Mathematics}

\author{Danilo T. Alves$^{1,2}$, Silvio C. F. Pereira Filho$^{2}$}
\affiliation{(1) - Faculdade de F\'\i sica, Universidade Federal do
Par\'a, 66075-110, Bel\'em, PA,  Brazil\\(2) - IEMCI, Universidade Federal do
Par\'a, 66075-110, Bel\'em, PA,  Brazil}

\date{\today}
\begin{abstract}

In the present paper, we describe some experiences in using programming, commands
and graphical interfaces based on computer algebra
systems as tools for learning Physics and Mathematics. 
\end{abstract}
%\pacs{.}

\maketitle
%

%\section{Introduction}

Algebraic computation is the manipulation of mathematical symbols 
made in the computer, according to the rules of mathematics. 
Some famous computer algebra systems are Maple \cite{maple}, Mathematica
\cite{mathematica} and Maxima \cite{maxima}.
These systems enable us to manipulate symbolic and numerical expressions, including integration, differentiation, matrices, and others. 
In the present paper, we describe an experience in using programming in Maple and Maxima 
as tools for learning
physics and mathematics. 

In the Federal University of Pará (Brazil), several pedagogical experiences using CAS as learning tools have
been developed \cite{Alves-Terrab-Medeiros-Amaral-RBEF-2002,alves-filho-souza-elias-RBEF-2010}. One of then was based in programming simple routines in Maple
as tools for learning electromagnetism \cite{Alves-Terrab-Medeiros-Amaral-RBEF-2002}. 
For instance, undergraduate students created, as a tool for learning, their own routines to calculate the operators 
{\it gradient},  {\it divergent} and {\it rotational}, named, respectively as 
{\tt grad}, {\tt div} and {\tt rot} \cite{Alves-Terrab-Medeiros-Amaral-RBEF-2002}:
\begin{verbatim}
> grad := proc(f) diff(f,x)+diff(f,y)+ diff(f,z) end:
> div := proc(v) diff(v[1], x)+diff(v[2], y)+diff(v[3], z); end:
> rot := proc(v)[diff(v[3], y)- diff(v[2], z),diff(v[1], z) 
 - diff(v[3], x), diff(v[2], x) - diff(v[1], y)]; end:
\end{verbatim}
The students could apply their own routines (operators) in scalar and vector fields, exploring themselves the
meaning of these operators. It was a successful experience, as described in Ref. \cite{Alves-Terrab-Medeiros-Amaral-RBEF-2002}.

Another pedagogical experience was done with students of public high schools. In this context, we used
Maxima instead of Maple, since the former is free. In this context, the students again used commands and
developed small routines for learning very introductory Physics. For instance, we show the
routine {\tt Conversor}, which was built to make transformations between multiples of meter:
\begin{verbatim}
Conversor(a,b,c) := 
if b=km then a*10^3*c 
else if b=hm then a*10^2*c 
else if b=dam then a*10*c 
else if b=m then a/1*c 
else if b=dm then float(a/10)*c 
else if b=cm then float(a/10^2)*c 
else if b=mm then float(a/10^3)*c;
\end{verbatim}
Application:
\begin{verbatim}
Conversor(3,km,m);
3000*m
\end{verbatim}
Again the students could apply their own routines, create questions, answer and test with the routines. 
It was again a successful experience. A problem observed in this application is in some difficulties arisen
by the syntax of Maxima. 
To contour this problem, we developed interactive graphical interfaces, free of syntax,
which are connected with a CAS. These interfaces (which can be done in PhP, Java, C, etc..) are intuitive,
enable students explore topics of Physics, Mathematics and also Chemistry,
solving analytical problems step-by-step, giving feedback about errors and showing interactive graphs.
In this kind of pedagogical application, the professor creates interactive interfaces for a certain issue,
working in the following manner.
Let us consider that $n$ students use the same number of computers (client computers).
When the student requires some calculation in the interface in his client computer,
the interface sends a process (an input) which: (a) can be processed by 
a CAS (as Maxima) installed in the own client computer; (b) can be processed in a CAS installed in a server machine,
in this case, remotely accessed (intranet or via internet). The output from the
CAS (for instance, Maxima) is sent to the interface in the client machine.
In the case of remote access, the server machine managers the queue of requested processes.

Another interesting situation, considering the current interest in electronic books (e-books),
is the use of a CAS to enable special interactivity into an e-book. Considering an e-book built in
some appropriate language (EPUB, for instance), the possibility of including
solution of analytical problems step-by-step, giving feedback about errors and showing interactive graphs,
could be reached attaching in the e-book file a CAS software. The e-book 
could send input lines to the CAS, and  gets the output. 
A javascript version of the CAS could be necessary to maintain the e-book compatible with the environment where it is usually run.

We believe that this paper can be useful for those interested
in using computer algebra as a pedagogical tool.

%%%%%%%%%%%%%%%%%%%%%%%%%%%%%%%%%%%%%%%%%%%%%%%%%%%%%%%%%%%%%%%%%%%%%%%%%%%%%%%

%%
\end{document}